\documentclass[conference]{IEEEtran}
\usepackage{eurosym}
\usepackage[ruled,vlined]{algorithm2e}
\usepackage{xcolor,soul,framed}
\usepackage[pdftex]{graphicx}
\usepackage[cmex10]{amsmath}
\usepackage{array}
\usepackage{mdwmath}
\usepackage{mdwtab}
\usepackage{eqparbox}
\usepackage{url}
\usepackage{flushend,cuted}
\usepackage[noend]{algpseudocode}
\usepackage{subfigure}
\usepackage{caption}
\usepackage{cite}
\usepackage{xcolor}
\usepackage{epstopdf}
\usepackage{cases}
\usepackage{amscd,textcomp,amssymb,epsfig,graphics,epsf,color,amsmath,balance,cite}
\usepackage{multirow}
\usepackage{booktabs}
\usepackage{stfloats}
\usepackage{setspace}
\usepackage{graphicx}
\usepackage{amsmath, amsthm, amsfonts, amssymb, amsbsy}
\usepackage{dsfont}
\usepackage{color}
\usepackage{bm}
\usepackage{tabu}
\usepackage{multirow}
\usepackage{multicol}
\usepackage{multirow}
\usepackage{float}
\usepackage{makecell}
\usepackage{booktabs}
\usepackage{flushend}
\usepackage{geometry}

\colorlet{shadecolor}{yellow}
\graphicspath{{../pdf/}{../jpeg/}}
\DeclareGraphicsExtensions{.pdf,.jpeg,.png}

\allowdisplaybreaks[4]
\begin{document}

\title{Parametric Near-Field MMSE Channel Estimation for sub-THz XL-MIMO Systems}
\author{\IEEEauthorblockN{Wen-Xuan Long${}^{*}$, Marco Moretti${}^*$, Michele Morelli${}^*$, Luca Sanguinetti${}^*$, Rui Chen${}^+$}
\IEEEauthorblockA{
${}^*$Dipartimento di Ingegneria dell’Informazione, University of Pisa, Italy \\
${}^+$State Key Laboratory of Integrated Service Networks, Xidian University, China \\
Email: wenxuan.long@ing.unipi.it*, marco.moretti@unipi.it, michele.morelli@unipi.it, luca.sanguinetti@unipi.it.}}
\maketitle

\begin{abstract}
Accurate channel estimation is essential for reliable communication in sub-THz extremely large (XL) MIMO systems. Deploying XL-MIMO in high-frequency bands not only increases the number of antennas, but also fundamentally alters channel propagation characteristics, placing the user equipments (UE) in the radiative near-field of the base station. This paper proposes a parametric estimation method using the multiple signal classification (MUSIC) algorithm to extract UE location data from uplink pilot signals. These parameters are used to reconstruct the spatial correlation matrix, followed by an approximation of the minimum mean square error (MMSE) channel estimator. Numerical results  show that the proposed method outperforms the least-squares (LS) estimator in terms of the normalized mean-square
error (NMSE), even without prior UE location knowledge. 
\end{abstract}

\section{Introduction}


Terahertz (THz) wireless communication is gaining increased interest in various applications (including space communications, biomedical sensing, and industrial automation) due to its potential advantages in terms of high bandwidth and low latency \cite{Chen2021Terahertz}. Several studies have investigated the transmission characteristics of wireless channels in the THz and sub-THz frequency bands \cite{Han2022Terahertz}. In \cite{bodet2024sub} and \cite{SELIMIS2024102453}, the authors measured and modeled the path loss in sub-THz bands, revealing that as the frequency increases, sub-THz channels experience greater propagation loss, leading to rapid signal power attenuation. Additionally, \cite{9135643} measured the power delay profile in the sub-THz band, showing that most line-of-sight (LoS) measurements lack significant multipath components, demonstrating the high directionality and narrow beamwidth of THz channels. Based on these findings, the THz wireless transmission is more likely to enable the short-range communication \cite{Okura2024Performance} by leveraging its strong LoS characteristics.

Considering the further expansion of array apertures in 6G base stations (BS), the user equipments (UEs) are likely to fall within the near-field (NF) region of the BS in THz short-range transmissions. In this region, the electromagnetic wavefront exhibits spherical curvature, introducing spherical phase variations across the array elements. These variations are determined by both the angle and distance between the array and the point source, fundamentally altering the channel propagation characteristics. Consequently, conventional far-field (FF) channel estimation methods \cite{10171211,10661241} become inadequate, necessitating the development of advanced NF channel estimation algorithms that fully exploit the unique properties of NF propagation.

A potential approach for the NF channel estimation is to first recover the
location parameters of the UE, and then incorporate them into the parametric
channel model to derive the channel estimate. In \cite{He2012Efficient} and 
\cite{Qu2024Two}, a two-stage multiple signal classification (MUSIC)
algorithm is proposed for uniform linear array (ULA)-based systems. In
this approach, the direction of arrival (DoA) of the UE is first estimated,
followed by the estimation of the distance between the UE and the BS using
the acquired angular information. Building upon this, \cite%
{Kosasih2023Parametric} extends the two-stage MUSIC algorithm to the NF
channel estimation based on uniform planar arrays (UPAs). The estimated
location parameters are then integrated into the NF channel model to derive
the channel estimates. However, the aforementioned studies primarily focus
on the parameter estimation for NF channels in low-frequency bands, while
the applicability of parametric channel estimation approaches in 
THz-band short-range communications is still unexplored.

In this paper, we propose a parametric channel estimation method for the sub-THz
NF channel based on the MUSIC algorithm. We assume that multiple UEs
periodically transmit orthogonal training sequences so as to avoid
multi-user interference at the BS. Furthermore, the signal transmitted by
each UE arrives at the BS through a line-of-sight (LoS) path with a
specified spatial spread. During channel estimation, the MUSIC algorithm is first applied to obtain
key parameters specifying the location of the UE relative to the BS. These
parameters are then used to reconstruct the spatial correlation matrix of
the NF channel. In doing so, we adopt mismatched values of the angular
and distance spreads, which are designed large enough to capture (almost
surely) all the received UE signal energy. The reconstructed channel
correlation matrix is eventually exploited to implement an approximation of the minimum
mean-square error (MMSE) estimator of the UE uplink channel.
Numerical simulations in the sub-THz band are conducted to evaluate the impact of transmission power, selected spatial spreads, and other parameters on the proposed solution. The results demonstrate that the proposed method
significantly outperforms the conventional least-squares (LS) estimator and other existing methods in
the sub-THz band.

\section{System Model}

We consider a communication system operating over a bandwith $B$ at sub-THz frequencies (e.g., in the range of $0.1$ THz) in which $K$
single-antenna UEs communicate with a BS equipped with an UPA, e.g., \cite[Fig. 1]{Bjornson2021Rayleigh}. The array is placed in the ${yz}-$plane of a three-dimensional
space, where a spherical coordinate system is defined, with $\varphi $ being
the azimuth angle, $\theta $ the elevation angle and $r$ the distance. The array has $N_H$
elements in each row and $N_{V}$ elements in each column, resulting in a
total of $N=N_H\times N_{V}$ elements. The horizontal and vertical inter-element spacing is $\delta$, while $\mathbf{v}_{n}$ is the location of $n$-th array element. 

The Fraunhofer distance of the array can be computed as $d_F = 2D^{2}/\lambda $ \cite{Cui2023Near},
where $D=\sqrt{(N_{V}^{2}+N_H^{2})}\delta$ is the array aperture length and 
$\lambda $ is the wavelength. This distance is used to distinguish between the far-field and radiative near-field regions of the array \cite{Cui2023Near}.
When the array aperture become large, reaching tens or even hundreds of times the wavelength, the expansion of the Fraunhofer distance makes it highly probable that the UEs would be located within the relative near-field region of the array. This is typically the case of sub-THz MIMO systems for short-range communications \cite{Okura2024Performance}. As an example, consider a half-wavelength-spaced array with size $0.15 \times 0.075$ m$^2$, and
operating at $0.1$ THz (i.e., $\lambda = 0.003$ m). This array contains
$5000$ antennas in the configuration $100 \times 50$, so that we have $d_F = 18.75$ m. In short-range communications (e.g., on the order of tens of meters), the UEs will likely
be located below $d_F$. This implies that the far-field approximation cannot be used, and the exact propagation model for the channel must be considered instead. This model is introduced next.


%
%

\subsection{Channel Model}
Based on \cite{bodet2024sub,SELIMIS2024102453,9135643}, a reasonable assumption in sub-THz short-range communications is that the signal transmitted by each UE arrives at the BS array within a small solid angle centered around the LoS path. We
concentrate on a given UE and denote by $( \varphi ,\theta )$ the LoS angles from
the UE to the array. Assuming a conventional correlated Rayleigh fading
channel, the channel vector $\mathbf{h} \in \mathbb{C}^N$ can be
modeled as \cite{Demir2024Spatial}
\begin{equation}
\mathbf{h}\sim \mathcal{N}_{\mathbb{C}}(\mathbf{0}_{N},\mathbf{R})
\end{equation}%
which is fully characterized by the spatial correlation matrix $\mathbf{R}$. It is further given by $ \mathbf{R} = \beta \mathbf{A}$, where $\beta =\frac{1}{N}\text{tr}\{\mathbf{R}\}$ is the average channel power (capturing
pathloss and shadowing), and $\mathbf{A}$ is expressed as
\begin{align}\label{RhNF}
&\mathbf{A}= \\ &\int_{ r-\triangle
_{r}}^{ r+\triangle _{r}}\!\!\!\!\int_{\varphi -\triangle
_{\varphi }}^{\varphi +\triangle _{\varphi }}\!\!\!\!\int_{ \theta
-\triangle _{\theta }}^{ \theta +\triangle _{\theta
}}\!\!f(\tilde r,\tilde \varphi ,\tilde \theta )  
\mathbf{a}(\tilde r,\tilde \varphi ,\tilde \theta )\mathbf{a}^{\mathrm{H}}(\tilde r,\tilde \varphi
,\tilde \theta )d\tilde \theta d\tilde\varphi d\tilde r,
\end{align}
where $%
(r,\varphi,\theta)$ are the distance, the azimuth and
elevation angles of the considered UE, while the triplet $(\triangle _{r},\triangle _{\varphi
},\triangle _{\theta })$ accounts for the corresponding distance and angular
spreads. Also, $\mathbf{a}(\tilde r,\tilde \varphi ,\tilde \theta )$ is the array response vector \cite{Liu2023Near}:%
\begin{equation}
\mathbf{a}(\tilde r,\tilde \varphi ,\tilde \theta )=\left[ 1,e^{j\frac{2\pi }{\lambda
}(\tilde r_{1}-\tilde r)},\ldots ,e^{j\frac{2\pi }{\lambda }(\tilde r_{N}-\tilde r)}\right] ^{%
\text{T}}  \label{array_response}
\end{equation}%
while $\tilde r_{n}$ and $\tilde r$ are the distances from the $n$-th array element and the
reference element to a point within the spatially spread region. The distance $\tilde r_{n}$ can be  computed as \cite{Liu2023Near}
\begin{align}
\tilde r_{n}= \tilde r \sqrt{1-\frac{2\mathbf{k}^{\text{T}}(\tilde \varphi ,\tilde \theta )\mathbf{v}_{n}%
}{\tilde r}+\frac{\Vert \mathbf{v}_{n}\Vert ^{2}}{\tilde r^{2}}} \label{rmU} 
\end{align}%
where $\mathbf{k}(\tilde \varphi ,\tilde \theta )=[\cos {\tilde\theta }\cos {\tilde\varphi },\cos {%
\tilde\theta }\sin {\tilde\varphi },\sin {\tilde\theta }]^{\text{T}}$ is the radiation
direction from the point to the array. Finally, $f(\cdot)$ is the normalized spatial
scattering function \cite{Demir2024Spatial}. %
Unlike in far-field conditions, the array
response vector in \eqref{array_response} is influenced not only by $\tilde\varphi 
$ and $\tilde\theta $, but also by the distance $\tilde r$. 


\subsection{Pilot Signal Model}

We assume a block fading channel model where the channel vectors remain static within a coherence block of $\tau_{c}$ channel uses. Within each block, $\tau_p\geq K$ uses are allocated for uplink channel estimation. We assume that the UEs transmit with power $p$ and are separated by means of \textit{orthogonal} pilot sequences of length $\tau_p$. Focusing on a generic UE $k$ and omitting the index for notational simplicity, the interference-free observation vector $\mathbf{y}\in \mathbb{C}^{N}$ is given by \cite{Bjornson2017Massive}:
\begin{equation}
\mathbf{y}=\mathbf{h}+\mathbf{w}  \label{n2.1}
\end{equation}%
where $\mathbf{w}\sim \mathcal{N}_{\mathbb{C}}(\mathbf{0}_{N},%
\sigma^2_w\mathbf{I}_{N})$ with $\sigma^2_w = N_0 B/(p\tau_p)$. The MMSE estimate of $\mathbf{h}$ is \cite{Bjornson2017Massive}%
\begin{equation}
\widehat{\mathbf{h}}^{\rm mmse}=\mathbf{R}\left(%
\mathbf{R}+\sigma^2_w\mathbf{I}_{N}\right)^{-1}\mathbf{y}.  \label{n2.23}
\end{equation}
Denoting by $\mu $ the
rank of $\mathbf{A}$, the eigenvalue
decomposition (EVD) of $\mathbf{A}$ yields
\begin{equation}
\mathbf{A}= \mathbf{U}\mathbf{\Lambda }\mathbf{U}%
^{\text{H}}  \label{A}
\end{equation}%
where $\mathbf{\Lambda }=\text{diag}\{\lambda _{1},\lambda _{2},\ldots
,\lambda _{\mu }\}$ collects the $\mu $
non-zero eigenvalues of $\mathbf{A}$, while $\mathbf{U}=[%
\mathbf{u}_{1}, \mathbf{u}_{2},\ldots,\mathbf{u}_{\mu }]^{T}$
collects the corresponding \textit{unit-norm} eigenvectors. Plugging \eqref{A} into \eqref{n2.23} yields the following equivalent expression:
\begin{equation}
\widehat{\mathbf{h}}^{\rm mmse}=\mathbf{U}\mathbf{\Lambda }%
\left(\mathbf{\Lambda } + \frac{\sigma_w^2}{\beta}\mathbf{\mathbf{I%
}}_{\mu }\right)^{-1}\mathbf{U}^{\text{H}} \mathbf{y},
\label{n2.4}
\end{equation}%
from which we see that the MMSE estimator requires knowledge of both $\mathbf{A}$ (or its EVD decomposition) and $\{\beta,\sigma^2_w\}$. This problem is addressed next.

\section{Parametric-based Channel Estimation}

From \eqref{RhNF}, we see
that, for a given array manifold and spatial scattering model, the spatial correlation matrix $\mathbf{R}%
$ is fully determined by the average channel power $\beta$, the UE location $(r,\varphi ,\theta )$ and the
triplet of intervals $(\triangle _{r},\triangle _{\varphi },\triangle
_{\theta })$. In practice, these parameters are not known. A possible solution consists of estimating the
location of the UE and fixing sufficiently large values of the
aforementioned intervals, say $(\bar{\triangle}_{r},\bar{\triangle}%
_{\varphi },\bar{\triangle}_{\theta })$, which allows us to design a 
\emph{refined} correlation matrix $\mathbf{\bar{R}}$.
Assuming $\bar{\triangle}_{r}>\triangle _{r}$, $\bar{\triangle}%
_{\varphi }>\triangle _{\varphi }$, $\bar{\triangle}_{\theta
}>\triangle _{\theta }$ and perfect estimation of $(r,\varphi
,\theta )$, from \cite[Lemma 1]{Demir2024Spatial} the subspace spanned by the columns of $\mathbf{\bar{R}}$ contains the subspace spanned by the columns of $\mathbf{R}$. Hence, the refined matrix $\mathbf{%
\bar{R}}$ contains all \textit{plausible} channel dimensions
and can replace $\mathbf{R}$ in the MMSE estimator \eqref{n2.4}. Next, the UE location $(r,\varphi
,\theta )$ is estimated by using the sample covariance matrix and the MUSIC algorithm.

\subsection{Parametric-based Reconstruction of $\mathbf R$}
We begin by  observing that $\mathbf{R}$ captures macroscopic effects such as spatial channel correlation and average path loss. Hence, it
changes slowly compared to $\mathbf{h}$ and maintains constant
over $\tau _{s}$ coherence blocks, where $\tau _{s}$ can be at the order of
hundreds or more \cite{Upadhya2018Covariance,Impact2021Kochar}. For this reason, we suppose that the BS has received the
pilot signal in (\ref{n2.1}) over $M\leq \tau _{s}$ coherence blocks and
denote by
\begin{equation}
\mathbf{y}(m)=\mathbf{h}(m)+\mathbf{w}(m),\ m=1,2,\ldots ,M, \label{num3}
\end{equation}%
the $M$ observations, with the channel and noise vectors 
being statistically independent for different values of \textit{m}. Notice that such observations can be obtained from pilots already used for
channel estimation in previous coherence blocks, so that no extra pilots are
needed. We use vectors $\mathbf{y}(m)$ to obtain the
sample correlation matrix as%
\begin{equation}
\widehat{\mathbf{R}}_{y}^{\rm sample}=\frac{1}{M}\sum_{m=1}^{M}\mathbf{y}(m)\mathbf{y}^{\mathrm{H}}(m)  \label{C}
\end{equation}%
and find a feasible method to get an estimate of parameters $(r, \varphi
,  \theta )$ from $\widehat{\mathbf{R}}_{y}^{\rm sample}$. Among several
existing schemes, the MUSIC is a powerful approach for estimating the
parameters of complex sinusoidal signals embedded in white Gaussian noise 
\cite{Schmidt1986Multiple}. The EVD of \eqref{C} can be
expressed as
\begin{equation}
\widehat{\mathbf{R}}_{y}^{\rm sample}=\sum_{n=1}^{N}\xi _{n}\mathbf{q}_{n}\mathbf{q}_{n}^{%
\mathrm{H}}=\mathbf{Q}_{y}\mathbf{\Sigma }_{y}\mathbf{Q}_{y}^{\mathrm{H}}
\label{R_y_1}
\end{equation}%
where $\{\xi _{n}|$$n$ $=$ $1,2,\ldots ,N\}$ are the eigenvalues of $\widehat{\mathbf{R}}_{y}^{\rm sample}$ and $\mathbf{q}_{n}\in \mathbb{C}^{N}$ is the
eigenvector corresponding to $\xi _{n}$. Furthermore, we have $\mathbf{%
\Sigma }_{y}$ $=$ $\text{diag}\{\xi _{1},$ $\xi _{2},\ldots ,\xi _{N}\}$ and 
$\mathbf{Q}_{y}=[\mathbf{q}_{1},\ldots ,\mathbf{q}_{N}]$. 
Since the channel and noise vectors in \eqref{num3} are independent, $\widehat{\mathbf{R}}_{y}^{\rm sample}$ can be decomposed 
as
\begin{equation}
\widehat{\mathbf{R}}_{y}^{\rm sample}=\mathbf{Q}^s_{y}\mathbf{\Sigma }^s_{y}(\mathbf{Q%
}^s_{y})^{\mathrm{H}}+\mathbf{Q}_{y}^{n}\mathbf{\Sigma }_{y}^{n}(\mathbf{Q}%
_{y}^{n})^{\mathrm{H}}  \label{EVD_Ry}
\end{equation}%
based on the rank $\widehat{\mu}_y = \text{rank}\{\widehat{\mathbf{R}}_{y}^{\rm sample}\}$\footnote{In the simulations, we utilize $\xi _{\widehat{\mu}_y}\geq10^{-2}$ to determine the rank of $\widehat{\mathbf{R}}_{y}^{\rm sample}$.}, where $\mathbf{\Sigma }^s_{y}\in \mathbb{C}^{\widehat{\mu}_y\times \widehat{\mu}_y}$ contains the $\widehat{\mu}_y$ largest eigenvalues of $\widehat{\mathbf{R}}%
_{y}^{\rm sample}$ and $\mathbf{Q}^s_{y}\in \mathbb{C}^{N\times \widehat{\mu}_y}$
spans the signal subspace of $\widehat{\mathbf{R}}_{y}^{\rm sample}$ formed by the
eigenvectors corresponding to these $\widehat{\mu}_y$ largest eigenvalues.
Furthermore, $\mathbf{\Sigma }_{y}^{n}\in \mathbb{C}^{(N-\widehat{\mu}_y)\times (N-\widehat{\mu}_y)}$ is the diagonal matrix composed of the
remaining eigenvalues and $\mathbf{Q}_{y}^{n}\in \mathbb{C}^{N\times (N-\widehat{\mu}_y)}$ is the noise subspace, spanned by the eigenvectors corresponding
to these smaller eigenvalues. Then, recalling that $\widehat{\mathbf{R}}_{y}^{\rm sample}$
converges (almost surely) to the true correlation matrix $\mathbf{R}_{y}=$E$%
\{\mathbf{y}(m)\mathbf{y}^{\mathrm{H}}(m)\}$
as $M\rightarrow \infty $, and that the channel vector $%
\mathbf{h}$ is orthogonal to the noise subspace
of $\mathbf{R}_{y}$, we construct the MUSIC metric as \cite{Schmidt1986Multiple}
\begin{equation}
P(\tilde{r},\tilde{\varphi},\tilde{\theta})=\frac{1}{\mathbf{a}%
^{\mathrm{H}}(\tilde{r},\tilde{\varphi},\tilde{\theta})\mathbf{Q}%
_{y}^{n}(\mathbf{Q}_{y}^{n})^{\mathrm{H}}\mathbf{a}(\tilde{r},\tilde{%
\varphi},\tilde{\theta})}  \label{Pmusic_NF}
\end{equation}%
where $\mathbf{a}(\tilde{r},\tilde{\varphi},\tilde{\theta})$
is the search vector with $\tilde{r}\in \lbrack 0,d_F]$, $\tilde{%
\varphi}\in \lbrack -\pi /2,\pi /2]$ and $\tilde{\theta}\in \lbrack -\pi
/2,\pi /2]$. 
The estimated UE location $(\hat{r}, \hat{\varphi}, \hat{\theta})$ is then obtained through the following
three-dimensional spectral peak search%
\begin{equation}
(\hat{r}, \hat{\varphi}, \hat{\theta})=\underset{(\tilde{r},\tilde{%
\varphi},\tilde{\theta})}{\text{arg}\max \{}P(\tilde{r},\tilde{%
\varphi},\tilde{\theta})\} \label{MUSIC}
\end{equation}
and used, together with the appropriately selected triplet $(\bar{\triangle}%
_{r},\bar{\triangle}_{\varphi },\bar{\triangle}_{\theta })$, to compute the
parametric-based correlation matrix as $\mathbf{\bar{R}}=\widehat{\beta}\mathbf{\bar{A}}$, where
\begin{align}\label{RhNF_bar} 
&\mathbf{\bar{A}}= \nonumber\\
&\int_{\hat{r}-\bar{\triangle}%
_{r}}^{\hat{r}+\bar{\triangle}_{r}}\!\!\!\!\int_{\hat{\varphi}-%
\bar{\triangle}_{\varphi }}^{\hat{\varphi}+\bar{\triangle}_{\varphi
}}\!\!\!\!\int_{\hat{\theta}-\bar{\triangle}_{\theta }}^{\hat{%
\theta}+\bar{\triangle}_{\theta }}\!\!f(\tilde r,\tilde \varphi,\tilde \theta)\mathbf{a}(\tilde r,\tilde\varphi ,\tilde\theta )\mathbf{a}^{\mathrm{H}}(\tilde r,\tilde\varphi
,\tilde \theta )d\tilde\theta d\tilde\varphi d\tilde r,
\end{align}%
and $\widehat{\beta}\!$ is an estimate of $\beta $, which is computed next.


\begin{table}[t!]
\renewcommand{\arraystretch}{1.3}
\setlength{\tabcolsep}{3pt}
\centering
\caption{System parameters.}
\begin{small}
\begin{tabular}{c|c}
\hline
{ \bf Parameter} & {\bf Value} \\
\hline
Carrier frequency & $f_0 = 0.1$\,THz \\
\hline
Wavelength  & $\lambda = 3$\,mm  \\
\hline
Number of antennas & $N_H \times N_V = 64 \times 32$ \\
\hline
Antenna spacing & $\delta= \lambda/{2}$ \\
\hline
Transmit power & $p = -4$ dBm \\
\hline
Pilot length & $\tau_p = 10$ \\
\hline
Number of observation vectors & $M = 10$ \\
\hline
Bandwidth & $B = 100$ MHz \\
\hline
Noise power & $N_0B = -84$ dBm \\
\hline
UE position & $(r,\!\varphi,\!\theta) = (4$ m $,\!-20^{\circ},\!-30^{\circ})$ \\
\hline
Average channel gain & $\beta = 90$ dB\\
\hline
Angular spread of elevation angle& $\Delta_{\theta} = 1.5^{\circ}$ \\
\hline
Assumed angular spread & $\bar \Delta_{\theta} = 5^{\circ}$ \\\hline
\end{tabular}
\end{small}
\label{TableI}
\end{table}

\subsection{Parametric-based Channel Estimator}

Letting $\bar{\mu}=\text{rank}\{\mathbf{\bar{R}}\} $, the EVD of the refined correlation matrix $\mathbf{\bar{R}}$ takes the form 
\begin{equation}
\mathbf{\bar{R}}=\widehat{\beta}\mathbf{\bar{U}}%
\mathbf{\bar{\Lambda}}\mathbf{\bar{U}} ^{%
\mathrm{H}} \label{R_NF_bar}
\end{equation}%
where $\mathbf{\bar{\Lambda}}$ $=$ $\text{diag}\{\lambda _{1},$ $%
\lambda _{2},\cdots ,\lambda _{\bar{\mu}}\}$ contains the $\bar{\mu}$
non-zero eigenvalues of $\mathbf{\bar{A}}$, and $\mathbf{%
\bar{U}}=[\mathbf{u}_{1},\cdots ,\mathbf{u}_{\bar{\mu}}]$ contains 
the
eigenvectors corresponding to these eigenvalues. As mentioned previously,
when $\bar{\triangle}_{r}-\triangle _{r}\geq \left\vert \widehat{r}%
-r\right\vert $, $\bar{\triangle}_{\varphi }-\triangle _{\varphi
}\geq \left\vert \widehat{\varphi}-\varphi \right\vert $ and $\bar{%
\triangle}_{\theta }-\triangle _{\theta }\geq | \widehat{\theta}%
-\theta | $, the subspace spanned by the columns of $%
\mathbf{\bar{U}}$ covers the subspace spanned by the columns of the
true $\mathbf{R}$. Hence, the parametric-based 
channel estimator based on $\mathbf{\bar{R}}$
takes the form 
\begin{equation}
\widehat{\mathbf{h}}^{\rm parametric}=\mathbf{\bar{U}}\mathbf{%
\bar{\Lambda}}\left( \mathbf{\bar{\Lambda}}+\frac{\widehat{\sigma}_w^2}{\widehat{\beta}\!}\mathbf{I}_{\bar{\mu}}\right) ^{-1}
\mathbf{\bar{U}}^{\mathrm{H}}\mathbf{y}
\label{MMSEest}
\end{equation}%
where $\widehat{\sigma}_w^2$ is an estimate of $\sigma_w^2$, which is computed next.

\subsection{Estimation of $\beta$ and $\sigma_w^2$}

We define $\mathbf{\bar{%
U}}^{n}\in \mathbb{C}^{N\times (N-\bar{\mu})}$ as the noise subspace
of $\mathbf{\bar{R}}$, spanned by the eigenvectors
corresponding to the $N-\bar{\mu}$ smaller eigenvalues. Then, we exploit the
observation vectors $\mathbf{y}(m)$ in (\ref{num3}) to
compute,
\begin{equation}
\mathbf{\bar{x}}_{n}(m)=\left( \mathbf{\bar{U}}^{n}\right) ^{\mathrm{H}}%
\mathbf{y}(m), \ m=1,2,\ldots ,M.
\label{vecx}
\end{equation}%
Observing that $\left( \mathbf{\bar{U}}^{n}\right) ^{\mathrm{H}}\mathbf{h%
}$ is ideally zero, after substituting (\ref{num3}) into (\ref{vecx})
we obtain $\mathbf{\bar{x}}_{n}(m)=\left( \mathbf{\bar{U}}^{n}\right) ^{%
\mathrm{H}}\mathbf{w}(m)=\mathbf{n}(m)$, where $\mathbf{n}(m)\sim \mathcal{N}%
_{\mathbb{C}}(\mathbf{0}_{N-\bar{\mu}},\sigma_w^2\mathbf{I}_{N-\bar{%
\mu}})$. An estimate of $\sigma_w^2$ is thus given by%
\begin{equation}
\widehat{\sigma}_w^2= \frac{1}{M(N-\bar{\mu})}\sum_{m=1}^{M}\Vert 
\mathbf{\bar{x}}_{n}(m)\Vert ^{2}.  \label{sigmaW}
\end{equation}%
Finally, observing that 
\begin{equation}
{\rm E}\left\{ \left\Vert \mathbf{y}(m)\right\Vert ^{2}\right\} =N\left(\beta +\sigma_w^2\right),
\end{equation}
an estimate of $\beta$ can be obtained as
\begin{equation}
\widehat{\beta}\!=\frac{1}{N}\text{tr}\{\widehat{\mathbf{R}}_{y}^{\rm sample}\}-\widehat{\sigma}_w^2,  \label{beta_est}
\end{equation}%
where $\widehat{\sigma}_w^2$ is given in (\ref{sigmaW}). The estimates $(%
\widehat{\beta},\widehat{\sigma}_w^2$) are eventually used in (\ref{MMSEest}) to evaluate $%
\widehat{\mathbf{h}}^{\rm parametric}$.

\begin{figure}[t]
\setlength{\abovecaptionskip}{-0.2cm} \setlength{\belowcaptionskip}{-0.4cm}
\par
\begin{center}
\includegraphics[width=8.5cm,height=5.5cm]{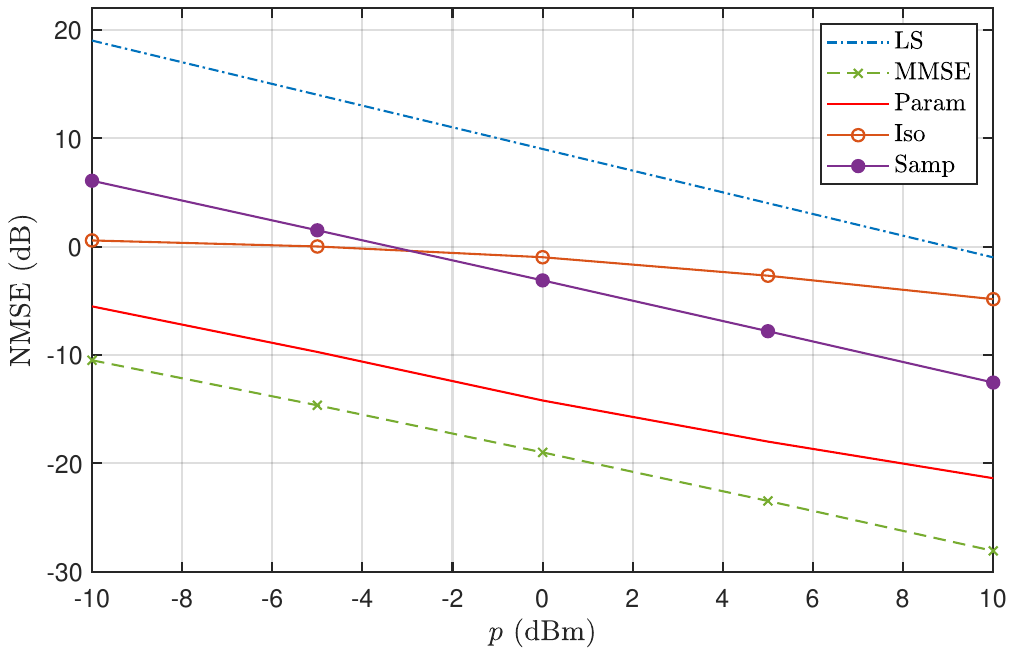}
\end{center}
\caption{The NMSE vs. $p$ for different estimators using the simulation
parameters provided in Table \protect\ref{TableI}.}
\label{Fig5}
\end{figure}
\begin{figure}[t!]
\setlength{\abovecaptionskip}{-0.2cm} \setlength{\belowcaptionskip}{-0.4cm}
\par
\begin{center}
\includegraphics[width=8.5cm,height=5.7cm]{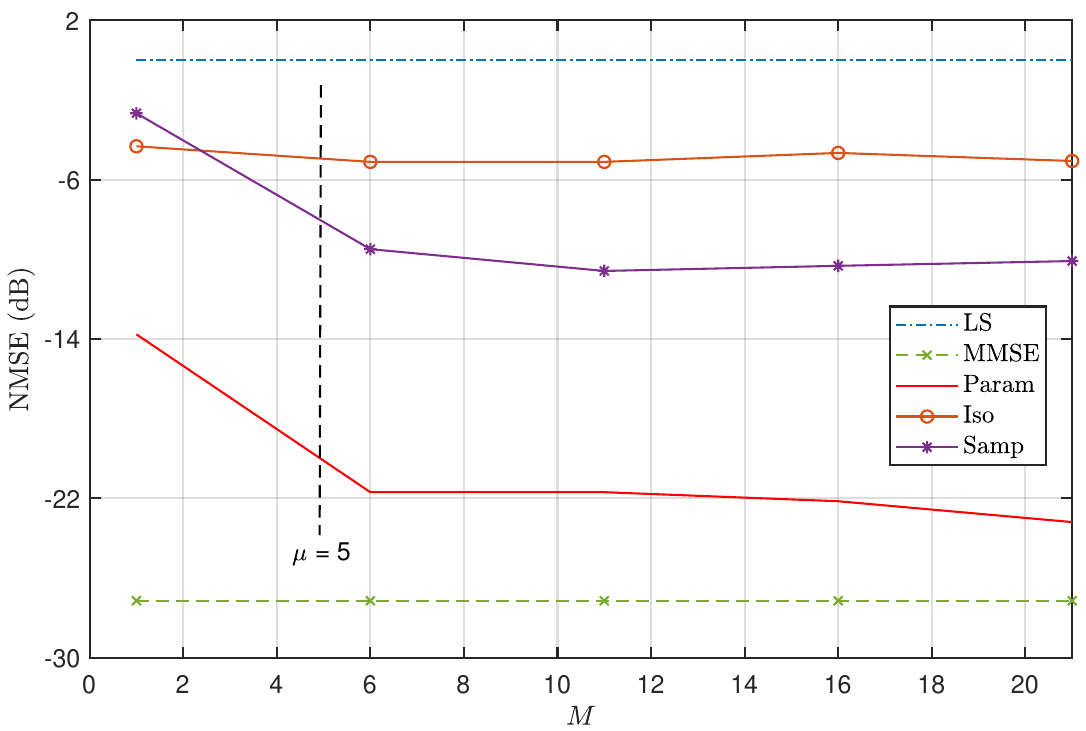}
\end{center}
\caption{The NMSE vs. $M$ for different estimators using the simulation
parameters provided in Table \protect\ref{TableI}.}
\label{Fig6}
\end{figure}

\section{Numerical Results}
We now evaluate the performance of the parametric channel estimation scheme \eqref{MMSEest} in terms of the normalized mean squared error (NMSE). To better reflect sub-THz channel performance, we adopt parameters consistent with existing studies: the path loss is set to $\beta = 90\ \rm{dB}$ \cite{SELIMIS2024102453}, the transmit power to $p = -4\ \rm{dBm}$ \cite{Yu2021A}, the uplink channel bandwidth $B = 100$ MHz \cite{Xing2022Sub} and the noise power spectral density to $N_0 = -174\ \rm{dBm}/Hz$ \cite{Okura2024Performance}.
The BS employs a UPA consisting of $64\times32$ elements, with an array aperture given by $D=\sqrt{%
(N_{V}^{2}+N_H^{2})}\delta=0.107$ m. This configuration corresponds to a Fraunhofer distance of $d_F = 7.680$ m. We consider a specific UE located at $(r, \varphi, \theta) = (4 \ \rm{m}, -20^{\circ}, -30^{\circ})$, which lies within the radiative near-field of the array.
In the MUSIC algorithm, we conduct a three-dimensional search over the space defined by $\tilde{r}\in[0.5\rm{m},\textit{d}
_{\textit{F}}], \tilde{\varphi}\in[-90^{\circ},90^{\circ}]$ and $\tilde{\theta}\in[-90^{\circ},90^{\circ}]$. The search is performed with the step sizes of $0.5\rm{m}$ for $\tilde{r}$ and $0.5^{\circ}$ for both $\tilde{\varphi}$ and $\tilde{\theta}$.
For the refined $\bar{\mathbf{R}}$, the elevation angle spread is set to $\bar \Delta_{\theta} = 5^{\circ}$, while the distance and azimuth spreads are computed using $\bar{\triangle}_{r}=\hat{r}(\cos (\hat{\theta}-%
\bar{\triangle}_{\theta })-\cos (\hat{\theta}+\bar{\triangle}%
_{\theta}))/2$ and $\bar{\triangle}_{\varphi}=\arctan (\bar{%
\triangle}_{r}/(\hat{r}\cos \hat{\theta}))$. Unless otherwise stated, other system parameters are listed in Table \ref{TableI}. In the simulation figures, we denote the parametric estimator derived from \eqref{MMSEest} as `Param'. For comparison, we evaluate its performance against the MMSE estimator \eqref{n2.4}, labeled as `MMSE', and three alternative methods:
\begin{itemize}
    \item The LS estimator: This method directly sets the channel estimate as
    $\widehat{\mathbf{h}}^{\rm ls}= \mathbf{y}$, labeled as `LS';
    \item The sample correlation matrix-based estimator, i.e.,
        \begin{equation}
\widehat{\mathbf{h}}^{\rm sample}=\left( \widehat{\mathbf{R}}_y^{\rm sample} - \widehat{\sigma}_w^2\mathbf{I}_N\right)\left(\widehat{\mathbf{R}}_y^{\rm sample}\right)^{-1}\mathbf{y},  \label{n2.3}
\end{equation}
which is labeled as `Samp';
    \item The isotropic correlation matrix-based estimator:
            \begin{equation}
\widehat{\mathbf{h}}^{\rm iso}=\widehat{\mathbf{R}}^{\rm iso}\left(%
\widehat{\mathbf{R}}^{\rm iso}+\widehat{\sigma}_w^2\mathbf{I}_N\right)^{-1}\mathbf{y}  \label{n2.3}
\end{equation}
with $\widehat{\mathbf{R}}^{\rm iso}$ being given by \cite[Eq. (8)]{Demir2024Spatial}. This estimator is labeled as `Iso'.
\end{itemize} 

\begin{figure}[t!]
\setlength{\abovecaptionskip}{-0.2cm} \setlength{\belowcaptionskip}{-0.4cm}
\par
\begin{center}
\includegraphics[width=8.5cm,height=5.7cm]{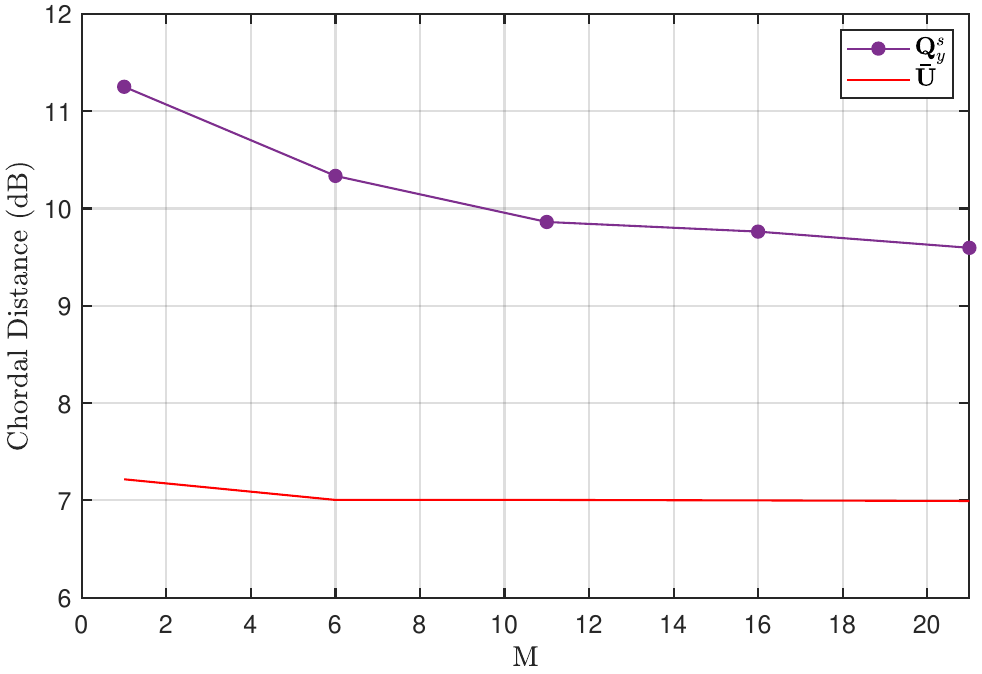}
\end{center}
\caption{The chordal distance vs. $M$ for different correlation matrices using the simulation parameters provided in Table \protect\ref{TableI}.}
\label{Fig10}
\end{figure}
\begin{figure}[t!]
\setlength{\abovecaptionskip}{-0.2cm} \setlength{\belowcaptionskip}{-0.4cm}
\par
\begin{center}
\includegraphics[width=8.5cm,height=5.7cm]{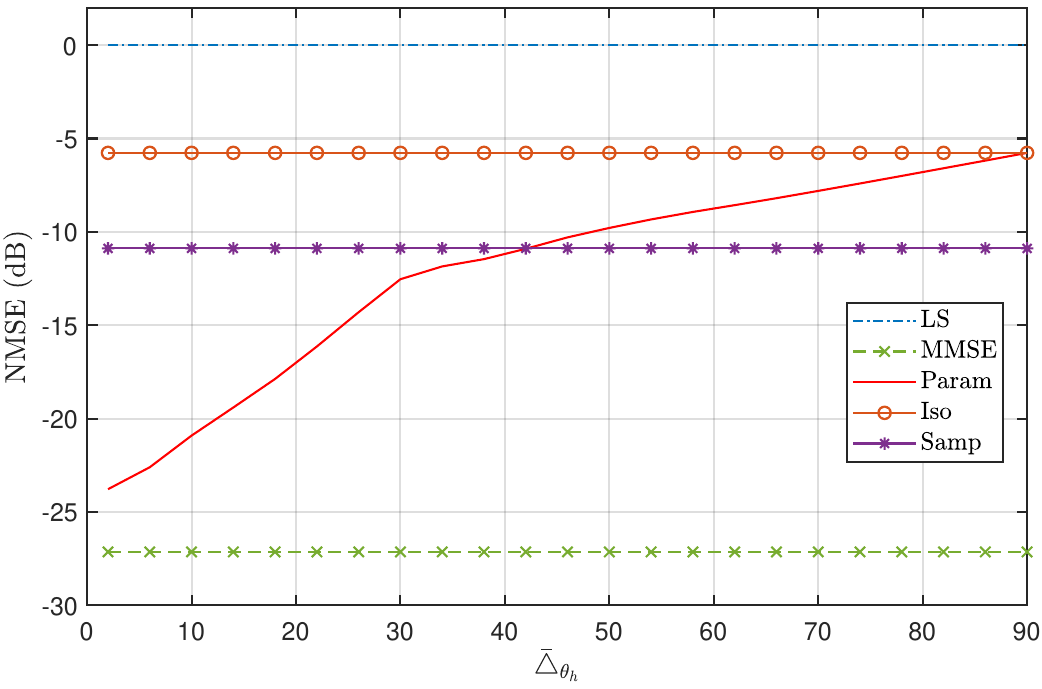}
\end{center}
\caption{The NMSE vs. $\bar{\triangle}_{\protect\theta}$ for different
estimators using the simulation parameters provided in Table \protect\ref{TableI}.}
\label{Fig7}
\end{figure}



Fig.~\ref{Fig5} plots the NMSE of the different estimators as a function of the transmit power $p$. As expected, the NMSE of all estimators decreases as the transmit power increases, reflecting the general trend that a higher SNR improves the channel estimation accuracy. More importantly, compared to existing schemes, the proposed estimator exhibits a significant performance advantage, ranking second only to the conventional MMSE estimator, which validates the effectiveness of the proposed method.

Fig.~\ref{Fig6} illustrates the NMSE as a function of the number of received pilot vectors, $M$. The figure also indicates $\mu=\text{rank}\{\mathbf{R}\}$. As observed, the proposed estimator is applicable for any $M\geq1$ and naturally improves as $M$ increases. The most significant NMSE reduction occurs when $M >\mu$, i.e., $M=6$, after which the performance gradually stabilizes, indicating that additional pilot vectors provide only marginal improvement. 

In Fig.\ref{Fig10}, we plot the chordal distance\footnote{The chordal distance $d_C(\mathbf{X},\mathbf{Y})$ between two matrices $\mathbf{X}$ and $\mathbf{Y}$ is defined as $d_C=\|\mathbf{X}\mathbf{X}^{\rm H}-\mathbf{Y}\mathbf{Y}^{\rm H}\|^2_F$} \cite[Eq. (7.21)]{Bjornson2017Massive} between the subspaces of the sample correlation matrix $\mathbf{Q}_y^s$ and the refined correlation matrix $\mathbf{\bar{U}}$ with respect to the actual subspace $\mathbf{U}$. 
From Fig.\ref{Fig10}, we observe that although the chordal distance of $\mathbf{Q}_y^s$ decreases as $M$ increases, it remains consistently higher than that of $\mathbf{\bar{U}}$. This further explains the performance advantage of `Param' over `Samp' as observed in Fig.~\ref{Fig6}.

Fig.~\ref{Fig7} evaluates the impact of the selected elevation angle spread $\bar{\triangle}%
_{\theta}$ on the proposed estimator for values ranging from $2^{\circ}$ to $90^{\circ}$. However, it is important to note that for sub-THz band channels, the angular spread is typically less than $10^{\circ}$. Here, we merely provide an example illustrating how the proposed estimator varies with $\bar{\triangle}%
_{\theta}$. As expected, the
performance of the proposed estimator gradually deteriorates as $\bar{%
\triangle}_{\theta}$ increases. Notably, when the selected angular spread approaches isotropy, i.e., $\bar{\triangle}_{\theta}\geq30^{\circ}$, the accuracy degradation gradually decreases. When the selected $\bar{%
\triangle}_{\theta} = 90^{\circ}$, i.e., the refined $\mathbf{\bar{R}}$ under the fully isotropic assumption, the proposed estimator demonstrates the same performance as the estimator in \cite{Demir2024Spatial}.



\section{Conclusions}

This paper considered the parametric near-field channel estimation problem in
the sub-THz band without any prior information. The MUSIC algorithm was used to
estimate the location of the UE relative to the BS. Using the estimated location, the spatial correlation matrix of the UE-BS channel was reconstructed and incorporated into an approximation of the MMSE estimator to derive the near-field channel estimate in the sub-THz band. Numerical results showed that the proposed method
significantly outperforms the conventional LS estimator and other existing methods. 
Given the computational complexity of 3-D MUSIC, our future work will focus on developing a low-complexity UE localization approach tailored for the THz band. This method will be integrated with a parametric channel estimation framework to not only reduce the computational overhead, but also to enhance the estimation performance.

\section*{Acknowledgment}
This work has been performed in the framework of the HORIZON-JU-SNS-2022 project TIMES, cofunded by the European Union. Views and opinions expressed are however those of the author(s) only and do not necessarily reflect those of the European Union.

\bibliographystyle{IEEEtran}
\bibliography{IEEEabrv,reference}

\end{document}